\documentclass[hidelinks,%
 reprint,
nofootinbib,
amsmath,amssymb,
aps,
pra,
]{revtex4-1}
\usepackage{graphicx}
\usepackage{dcolumn}
\usepackage{bm}
\usepackage[makeroom]{cancel} 
\usepackage{ragged2e} 
\usepackage{changepage} 
\usepackage{tabularx} 
\usepackage{booktabs} 
\usepackage{todonotes} 
\usepackage{hyperref}

\usepackage [autostyle, english = american]{csquotes}
\MakeOuterQuote{"}

\usepackage{xcolor} 

\begin{document}

\preprint{APS/123-QED}

\title{Making large overhangs in micrometer and nanometer-sized structures}

\author{N. Van Horne}
\email{noah.vanhorne@protonmail.com}
\affiliation{Centre for Quantum Technologies, National University of Singapore, Singapore 117543}
\author{M. Mukherjee}
\affiliation{Centre for Quantum Technologies, National University of Singapore, Singapore 117543}
\affiliation{MajuLab, CNRS-UNS-NUS-NTU International Joint Research Unit, UMI 3654, Singapore}

\date{\today}

\begin{abstract}	
	We describe two general procedures for fabricating microstructures with large overhangs and high aspect-ratio support pillars. The first method uses a static angled dry etch on micro- or nano-pillars to create an initial overhang, followed by wet etching for further erosion. The second method uses a time-dependent angled etch on a flat plane patterned with protective resin, to reduce the number of lithography steps needed to make these objects. The time-dependent dry etch is again followed by a wet etch. For the second method we derive a formula that provides the rate at which the attack angle must evolve, given a known etch rate within the target material, the depth of the desired overhang (undercut), and the instantaneous attack angle.
\end{abstract}



\maketitle


			Creating an overhang in structures on the micrometer scale has applications ranging from making micro-disk resonators for use in photonics \cite{wang2012high}, to improved adhesion between layers of different materials \cite{larsson2005improved}, and producing large electrodes atop narrow, low-capacitance conducting pillars for potential quantum computing architectures \cite{CouplingDistantIons}. Existing techniques to produce overhangs primarily use wet etching. Wet etching may be isotropic, producing scallops along the edges of a vertical surface, or anisotropic, if etching proceeds through a sacrificial layer \cite{stonas2001photoelectrochemical} or if it progresses preferentially along certain axes of the crystalline lattice of the target material \cite{wang2012high}. Here, we describe two alternative methods to create overhangs using a combination of angled dry etching \cite{cybart2013nanometer}, and wet etching, to produce a wide plateau which sits upon a narrow support. Since both methods make use of the shadow cast by a protective mask, they can be considered two variants of shadow etching.
			
			\section{Method 1: Starting from a pre-fabricated structure}\label{PreFabMicroStruct}
			
\begin{figure}[h]
				\centering
				\includegraphics[width = 0.5\textwidth]{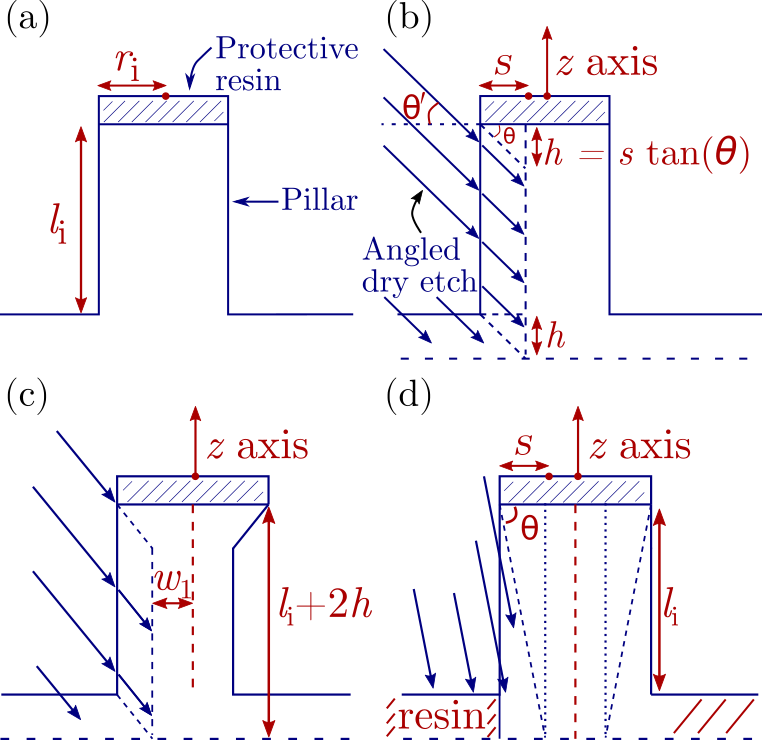}
				\caption{Starting pillar (a) followed by angled dry etching (b)-(d). (a) A structure is built with a protective resin covering. (b) An angled dry etch moves towards the center of the structure, progressing at an angle $\theta$ from the plane of the sample. (c) The sample is rotated by $180^\mathrm{o}$ around the $z$ axis to produce a symmetrical result. (d) If the substrate in which the pillar is mounted is protected by resin, and if the penetration angle $\theta$ is large, the pillar becomes tapered from top to bottom.}
				\label{Step1Step2}
			\end{figure}

			The process can be summarized as three steps:
			\begin{enumerate}	
				\item Production of pillars (or other objects)
				
				\item Angled dry etching of pillars (or other objects)
				
				\item Wet etching to further reduce the size of the support structure
			\end{enumerate}
			The first step may be implemented by a variety of methods ranging from top-down etching, for example using ion milling or DRIE, to building a structure from the bottom-up by repetitive stacking of squares or other shapes. Here, it will be assumed that the starting structures are square pillars. At the end of the first step, a protective resin should cover the structure to protect the top from a dry etch, as shown in Fig. \ref{Step1Step2} (a).
			
			In the second step, dry ballistic reactive ion etching (RIE) is performed with an attack angle $\theta '$ between the plane of the substrate and the trajectory of the etching agent (Fig. \ref{Step1Step2} (b)). Within the substrate, etching proceeds at an angle $\theta$, which may be equal to $\theta '$ if the dry etch rate is isotropic, but can be different from $\theta '$ in general. If the dry etch is anisotropic, the angle-dependent relationship between the attack angle $\theta '$ and the penetration angle $\theta$ is often known, $\theta ' = f \left( \theta \right)$, and each angle can be expressed in terms of the other. Therefore, for the remainder of this paper we usually reason only in terms of the penetration angle $\theta$. During the angled dry etch, the mask at the top of the pillar casts a shadow of vertical height $h = s\tan{\theta}$, where $s$ is the distance from the edge of the mask which determines the initial overhang. The rest of the vertical height of the pillar is exposed, and etching proceeds inwards as a uniform front along the angle $\theta$.
			
			When the etch is complete, the level of the substrate on which the pillar is mounted is also reduced by a distance $h = s\tan{\theta}$, assuming the pillar is made of the same material as the substrate in which it is mounted. This increases the effective height of the pillar relative to the surrounding surface by $h$, as shown in Fig. \ref{Step1Step2} (b). For a symmetrical result, the sample is then rotated by $180^\mathrm{o}$ around the $z$ axis, and the angled etch is repeated to create a second overhang (Fig. \ref{Step1Step2} (c)). To make overhangs on the remaining two sides of the square pillars, the angled etch must be repeated two more times, at rotations of $90^{\mathrm{o}}$ and $270^{\mathrm{o}}$, resulting in a square support column with $\sim 90^{\mathrm{o}}$ angles between the sides of the column and the plane of the substrate. We note that a symmetrical result might also be accomplished by placing the sample on a tilted stage that rotates continuously. From an experimental standpoint this could be a simpler fabrication process. However, this approach poses a theoretical challenge; calculating the total depth of the etch in the pillar during a slow rotation is not simple, and it is difficult to make quantitative predictions. The calculation can be simplified by using a high rotation speed compared to the rate of inward etching, so that pillars are etched down evenly on all sides.\footnote{The rigorous criterion to satisfy is that the depth etched during half of a rotation is small compared to the desired precision. For dry etch rates on the order of $\sim 100~$ nm per minute, a precision on the order of $100~$nm can be achieved for rotation periods as long as $2$ minutes per cycle, without compromising the total process time (i.e. using high dry etch rates).} Such an approach is left for a separate theoretical study, and the remainder of this article assumes that angled etching is performed in $4$ steps, each at different fixed rotations around the $z$ axis. Before moving to other technical considerations, we observe that if the plane in which the pillar is mounted is protected by resin, in the event that the penetration angle $\theta$ is large such that the height $h$ of the shadow cast by the mask is greater or equal to the initial height $l_i$ of the support, or in other words if $\theta \ge \tan^{-1}{\left(l_i / s\right)}$, the support becomes tapered from top to bottom (Fig. \ref{Step1Step2} (d)).
			
\begin{figure}[h!]
				\centering
				\includegraphics[width = 0.4\textwidth]{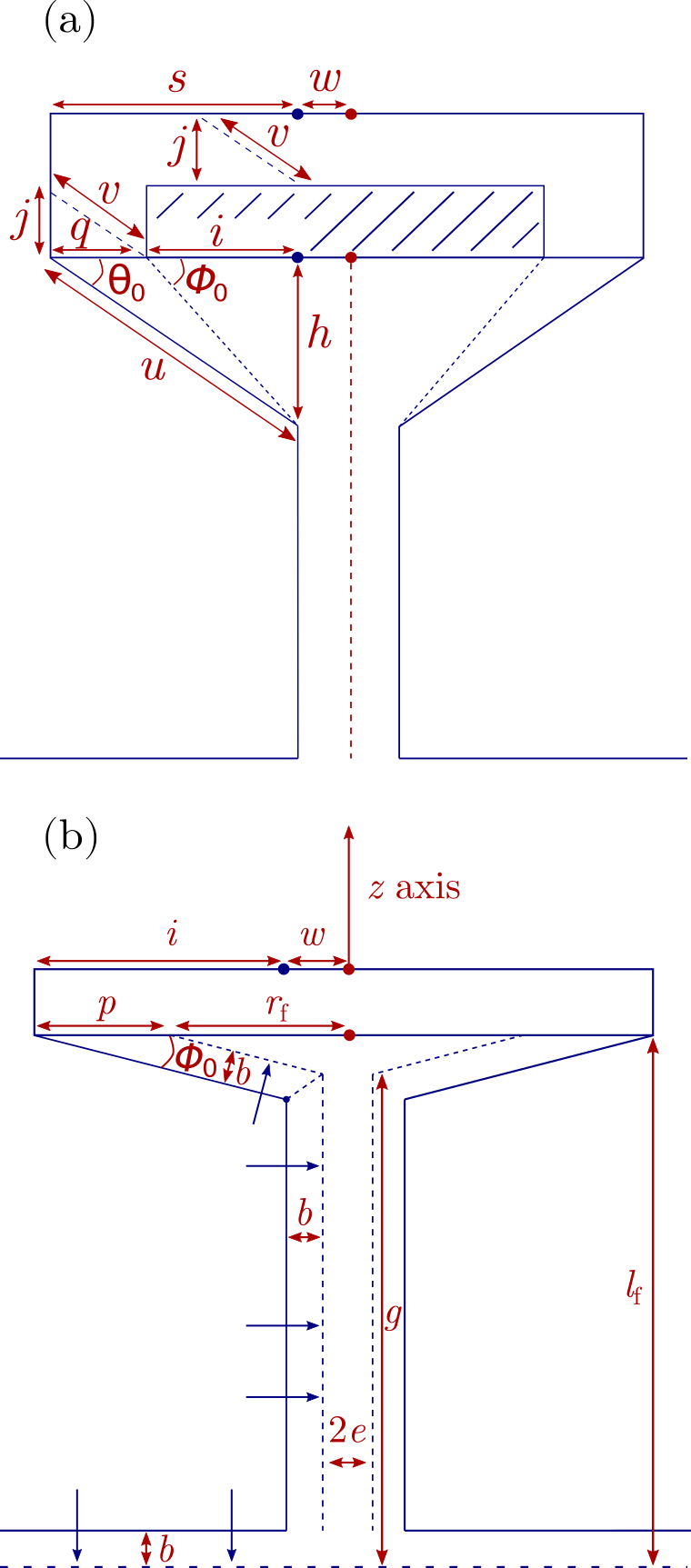}
				\caption{Effects of the dry etch on the protective resin (a), and wet etching (b). (a) In real applications, the angled dry etch removes some of the protective resin covering the top plateau. The portion $q$ of the top plateau which becomes exposed depends on the selectivity of the etch, $k = u/v$. (b) After the angled dry etches are complete, wet etching further reduces the diameter of the pillar while relieving mechanical stress symmetrically. If the wet etch is isotropic, a layer of thickness $b$ is removed from all exposed areas of the pillar. The final half side-length of the upside-down pyramid on top of the pillar is $r_\mathrm{f}$. The final diameter of the pillar is $2e$ and the final height of the support pillar is $g$.}
				\label{DryAndWetEtch}
			\end{figure}

			In real applications, the protective resin on the top plateau is also susceptible to the angled dry etch, and the amount it is etched depends on the selectivity ratio, $k = \frac{\left(\mathrm{distance~ etched~ in~ target~ material}\right)}{\left(\mathrm{distance~ etched~ in~ resin}\right)}$. Removal of the resin leads to a reduction in the size of the top plateau, as shown in Fig. \ref{DryAndWetEtch} (a).\footnote{Figure \ref{DryAndWetEtch} and all calculations from this point forward rely on the fairly strong assumption that if there is anisotropy in the dry etch rate, either within the target substrate or the protective resin, the anisotropy of the etch rate in both materials is identical. The validity and impact of this assumption should be considered on a case-by-case basis.} Figure \ref{DryAndWetEtch} (a) shows that the maximum distance $v$ etched in the resin is related to the maximum distance etched in the pillar, $u = s/\cos{\theta_0}$. Here, we have changed the notation for the angle of penetration from "$\theta$" to "$\theta_0$" to highlight that $\theta_0$ refers specifically to the initial angle of penetration. The relationship between $v$ and $u$ is $v=u/k$. A triangle of resin with hypotenuse $v$ is etched away, exposing a distance $q = v\cos{\theta_0}$ of the top plateau to the angled etch. Combining these relations, $q = \left(\left(s/\cos{\theta_0}\right)\big/k\right)\cos{\theta_0} = s/k$. The half side-length $r'$ of the top platform after performing the dry etch is then:
			\begin{equation}\label{r'}
			r' = (s - q) + w = s \left( 1 - 1/k \right) + w =i+w.
			\end{equation}
			The dimension $r'$ corresponds to $p + r_\mathrm{f}$ in Fig. \ref{DryAndWetEtch} (b). 
			
			The third step of wet etching allows mechanical stress to be relieved symmetrically from the support, preventing potentially destructive warping as its fragility increases. For example, suppose the four angled dry etches at $90^\mathrm{o}$ from each other reduce the support column to a square pillar $20~\mu$m thick. The wet etch is then used to erode material isotropically and reduce the pillar to a thickness of $5~\mu$m. During the wet etch, all parts of the structure including the top plateau are etched, as shown in Fig. \ref{DryAndWetEtch} (b). 
		
			We now illustrate an explicit example following figures \ref{Step1Step2} and \ref{DryAndWetEtch}. Before proceeding it is worth mentioning that numbers used in the examples herein are for "large" structures, with dimensions on the order of $1$ to $300~\mu$m. These sizes correspond to optimal dimensions for potential quantum computing applications, where ions can only be trapped at distances greater than $\sim 50$ micrometers from a room-temperature surface, due to parasitic heating. From a practical standpoint, these large dimensions mean that a real implementation of this example is likely to entail lengthy angled dry etches. We do not presently know typical dry etch rates for the superconducting materials that would constitute the pillars used in quantum computing applications. However, etch rates of conducting metals such as platinum, gold, and silver can be used as a rough guideline. The dry etch rate of Pt using Ar-O2 chemically assisted ion-beam etching (CA-IBE) at $700~$eV and normal incidence with a beam current of $1100~$mA [sic] is $\sim 130~$ nanometers per minute~ \cite{gutsche2000}, while the dry etch rate of Au using ion beam etching (IBE) with argon at $500~$eV at normal incidence and a current density of $1~$mA/cm$^2$ is ($\sim 108~$nm/min) \cite{EtchRates}, and the dry etch rate of Ag using the same conditions is ($\sim 180~$nm/min) \cite{EtchRates}. These figures, on the order of $6~\mu$m/hr, suggest that etching a distance of $\sim 100~ \mu$m could require up to $20~$hrs. Using RIE in place of IBE might reduce this time. Since it is useful in the example below to have a value for the selectivity $k$, we mention that the dry etch selectivity of Pt compared to Ti using Ar-O2 chemically assisted ion-beam etching at $700~$eV and normal incidence with a beam current of $1100~$mA [sic] is $k \sim 22~$, \cite{gutsche2000} although the selectivity decreases for angles $\theta < 90^{\mathrm{o}}$. For a rough estimate of the time needed for the wet etch, one can use the guideline of Ag etched by a piranha etch, which can progress at up to $600~$nm/min.~\cite{williams2003etch} This suggests that the erosion of $\sim 10 ~\mu$m of the support pillar could require on the order of $20$ minutes. Although the time-frame of the aforementioned angled etch may seem impractical, we stress that this is not a limitation of the method. It is a particularity of the example considered below, geared towards quantum computing applications. The principles and equations discussed herein are, in general, equally valid for creating smaller structures on the nanometer scale.
			
			In this example, first a square pillar is manufactured with side length $2 \times r_\mathrm{i} = 2 \times 103~\mu$m and height $l_\mathrm{i} = 150~\mu$m (Fig. \ref{Step1Step2} (a)). Second, dry etching is performed with an angle $\theta_0 = 5^{\mathrm{o}}$, inwards a horizontal distance $s = 93~\mu$m. The height of the pillar is increased by $h = s\tan{\theta_0} \sim 8~\mu$m. The sample is rotated $180^{\mathrm{o}}$ around the $z$ axis for a second etch, and again the height of the pillar is increased by $h$. This results in a wall $20~\mu$m thick, $206~\mu$m long, and $\sim 166~\mu$m tall. The sample is then rotated to angles of $90^{\mathrm{o}}$, and $270^{\mathrm{o}}$ around the $z$ axis, and etched a third and fourth time. Overall, the angled etches increase the height of the pillar by $4h$. If we label $w_1$ the distance from the center of one side of the support pillar, to the center of the pillar after the first angled etch, (and we give the same label to the distance from the edge of the support pillar to the center of the pillar after the second angled etch, at a rotation of $180^{\mathrm{o}}$), and we label $w_2$ the distance from the center of one side of the support pillar to the center of the pillar after the etches at $90^{\mathrm{o}}$, and $270^{\mathrm{o}}$, the dimensions of the pillar are then $2w_1 \times 2w_2 \times \left( l_\mathrm{i} + 4h \right) \sim 20\times20\times182~\mu\mathrm{m}^3$. Third, isotropic wet etching erodes inwards, upwards and downwards a distance $b=7.5~\mu$m from each side (Fig. \ref{DryAndWetEtch} (b)). This reduces the lateral dimensions of the support pillar to a cylinder-like column with characteristic "radius" $e = w - b = 2.5~\mu$m. (The term "radius" is used because wet etches are known to round off sharp edges such as the $90^{\mathrm{o}}$ corners of the support pillar)\cite{williams2003etch}. The wet etch also removes a region which can be approximated as a trapezoidal prism from the top plateau, with inner angle $\phi_0$ and height $b \sim 7.5~\mu$m, as shown in Fig. \ref{DryAndWetEtch} (b). We recall that for non-infinite etching selectivities, the angle $\phi_0$ is not equal to $\theta_0$ (see Fig. \ref{DryAndWetEtch} (a)). The general relationship between $\phi_0$ and $\theta_0$ is found by writing $\phi_0 = \tan^{-1}{\left(h/i\right)}$. Plugging in $i = s \left( 1 - 1/k \right) $ from equation \eqref{r'}, and $h = s \tan{\theta_0}$ from figure \ref{Step1Step2} (b), gives $\phi_0 = \tan^{-1}{\left( \tan{\theta_0}/\left( 1-1/k \right) \right)}$. Overall, the angled etch and the wet etch reduce the half-width of the top plateau as follows: first, the angled dry etch reduces the half side-length from $r_\mathrm{i} = s + w$, down to $r' = s \left( 1 - 1/k \right) + w = i +w$ (equation \eqref{r'} and Fig. \ref{DryAndWetEtch} (a)). Then, the wet etch brings it to a final value $r_\mathrm{f} = \left(r'-p\right)$, where $p = b / \sin{\phi_0}$ (Fig. \ref{DryAndWetEtch} (b)). Combining the two reductions of the top plateau, for the given example and assuming a selectivity $k = \infty$ leads to: 
			\begin{equation}\label{rf}
			r_\mathrm{f} = s \left( 1 - 1/k \right) + (e + b) - b / \sin{\phi_0}  \sim 17~\mu \mathrm{m}.
			\end{equation}
			Additionally, the wet etching erodes the plane in which the pillar stands by $b=7.5~\mu$m, increasing the height to $l_\mathrm{f}=l_\mathrm{i}+4h+b \sim 189.5~\mu$m.  Thus, the final result is an upside-down square pyramid of base-dimension $r_\mathrm{f} = 17~\mu$m and inner angle $\phi_0 = \theta_0 = 5^{\mathrm{o}}$ perched on a support pillar of radius $e = 2.5~\mu$m and height $g \approx \left( l_\mathrm{f} - r_\mathrm{f} \tan{\phi_0} \right) \sim 188~\mu$m.
		
			We can make a few observations about expression \eqref{rf}. First, the distance $s$ is linked to the etching angle by $s = u \cos{\theta_0}$, where $u$ is the total depth of material removed by the angled dry etch (Fig. \ref{DryAndWetEtch} (a)). This ensures that in the limiting case $\theta_0 \rightarrow 90^\mathrm{o}$ (and still assuming infinite selectivity such that $\phi_0 = \theta_0$), $r_\mathrm{f}$ tends to the correct value $r_\mathrm{f} = e$. In the opposite limiting case $\theta_0 \rightarrow 0^\mathrm{o}$, $r_\mathrm{f}$ becomes negative which is non-physical. This happens because as the angle $\theta_0$ decreases, if the wet etching depth $b$ remains fixed, eventually $r_\mathrm{f}$ reaches zero. When the depth $b$ of the wet etch exceeds the available material, expression \eqref{rf} becomes meaningless. Second, for practical applications one would generally select target values for $r_\mathrm{f}$, $e$, and $\phi_0$ and may also be restricted in the choice of protective resin, which imposes $k$. This leaves two free parameters, $s$ (   or $u \cos{\theta_0}$, or $u \cos{ \big( \tan^{-1} \left( (1-1/k) \tan ( \phi_0 ) \right) \big) }$   ), and $b$. Therefore, equation \eqref{rf} is more useful when rewritten in the form: 
			\begin{eqnarray}\label{CalcConvenient}
			&& u = \left[b \left(\frac{1}{\sin{\phi_0}} - 1 \right) + r_\mathrm{f} - e \right] 
			\nonumber \\
			&& \times \frac{1}{\cos \left(
			\tan^{-1} \left( \left( 1 - 1/k \right) \tan \phi_0 \right) \right)
			\left( 1 - 1/k \right)}.
			\end{eqnarray}
			If one sets out to achieve a top plateau of half side-length $r_\mathrm{f} = 50 ~\mu$m, a support pillar of radius $e = 2.5~\mu$m, an internal angle $\phi_0 = 5^{\mathrm{o}}$, and using a protective resin with selectivity $k = 20$, the values of $b$ and $u$ that satisfy these constraints can be plotted as a straight line with all of the positive (in other words, meaningful) values of $b$ and $u$ in the first quadrant. Note that fixing $\phi_0$ and the selectivity $k = 20$ fixes the penetration angle $\theta_0$ of the dry etch, since $\theta_0 = \tan^{-1}\left( \left(1-1/k\right) \tan{\phi_0} \right) = 4.76^\mathrm{o}$. One would typically decide on a target dimension $w = e+b$ achieved by the angled dry etch such that the support pillar remains sturdy enough to be carved reliably without risk of destruction. For instance, $w = 10~\mu$m could be adequate. Based on this, $b = w - e$, and in turn the dimension $u$ becomes fixed. Given the dimension $u$ and the etching angle $\theta_0$, the side length of the initial square pillar is fixed as $2 r_\mathrm{i} = 2 \times \left( u \cos{\theta_0} + w \right)$, and one can proceed with pillar manufacturing. For the example above with $w = 10~\mu$m, we find $b = 7.5~\mu$m which leads to $u \approx 114 ~\mu$m and hence a side length for the initial square pillar of $2r_\mathrm{i} = 2 \times \left( u \cos{4.76^{\mathrm{o}}} + 10 \right)~\mu \mathrm{m} \approx 247 ~\mu$m.
		
			Equation \eqref{CalcConvenient} can be evaluated to find $u$ by plugging in numbers, or the value of $u$ can be found by reading off the y-axis on the straight line plot of $u$ as a function of $b$. We recommend using a straight line plot, as not all choices of input parameters yield physically possible results. If the constraints cannot be satisfied, solving expression \eqref{CalcConvenient} may yield a negative result. A straight line plot immediately allows one to identify combinations where $b$ and $u$ are both positive. Even for cases where both $b$ and $u$ are positive, however, an additional constraint must be considered. It was mentioned previously that when the depth $b$ of the wet etch exceeds the available material, expression \eqref{rf} becomes meaningless. This condition can be expressed quantitatively as an upper bound on $b$. Assuming infinite selectivity, the depth of the wet etch is equal to the available material when $\tan{\theta_0} = b/u$ (see Fig \ref{DryAndWetEtch} (b) and (a)). Thus, to ensure a physically possible situation, the ratio of the distance $\sim h$ to the distance $u$ (Fig. \ref{DryAndWetEtch} (a)) must be larger than the limiting ratio. The condition to satisfy is $\tan{\theta_0 \ge b/u}$, or $b \le u \tan{\theta_0}$. When the selectivity is not infinite, the bound is given by the criterion $\tan{\phi_0} \ge b/c$, or $b \le c \tan{\phi_0}$, where $c$ is defined to be the length of the dashed line which defines the angle $\phi_0$ in Fig. \ref{DryAndWetEtch} (a) (not labeled). In practice, these bounds serve mostly as conceptual guidelines. For real implementations there should be no need to approach the upper bound on $b$, and in general it is advisable to remain as far from it as possible (in other words, to make $b$ as small as possible), while still respecting the minimum-width constraint $w \gtrsim 10~\mu$m, which sets the relationship $b = w - e$. Nevertheless, it is useful to calculate the upper bound on $b$ if for no other reason than to verify that one is well below it.
		
			For practical implementations, it is also advisable to use angles $\phi_0$ which are larger than the $5^{\mathrm{o}}$ given in the examples above. This leads to a more reliable result, as a larger angle mitigates error which arises due to imprecise timing either during the dry etch, or during the reduction of the top plateau by the wet etch. Referring to Fig. \ref{DryAndWetEtch} (b), for an internal angle $\phi_0 = 5^{\mathrm{o}}$ of the upside-down pyramid, and a final target half side-length $r_\mathrm{f} = 50~\mu$m, the height of the \textit{tallest} part of the upside-down pyramid, is $r_{\mathrm{f}} \tan{\phi_0} = 4.4~\mu \mathrm{m}$. For a top plateau this thin, if wet etching proceeds at a rate of several hundred nanometers per minute, errors in the timing of the wet etch could lead to completely dissolving the top plateau or making it larger than theoretically anticipated. In general, if $E_R$ is the "raw error" due to imprecision in the timing of the wet etch, the "leveraged error" $E_L$ in the final target half side-length of the upside-down pyramid is related to $E_R$ and the angle $\phi_0$ according to $E_L = E_R / \sin{\phi_0}$.

			\section{Method 2: Starting from a planar surface}\label{StartPlaneSurf}
			
\begin{figure}[h]
			\centering
			\includegraphics[width = 0.5\textwidth]{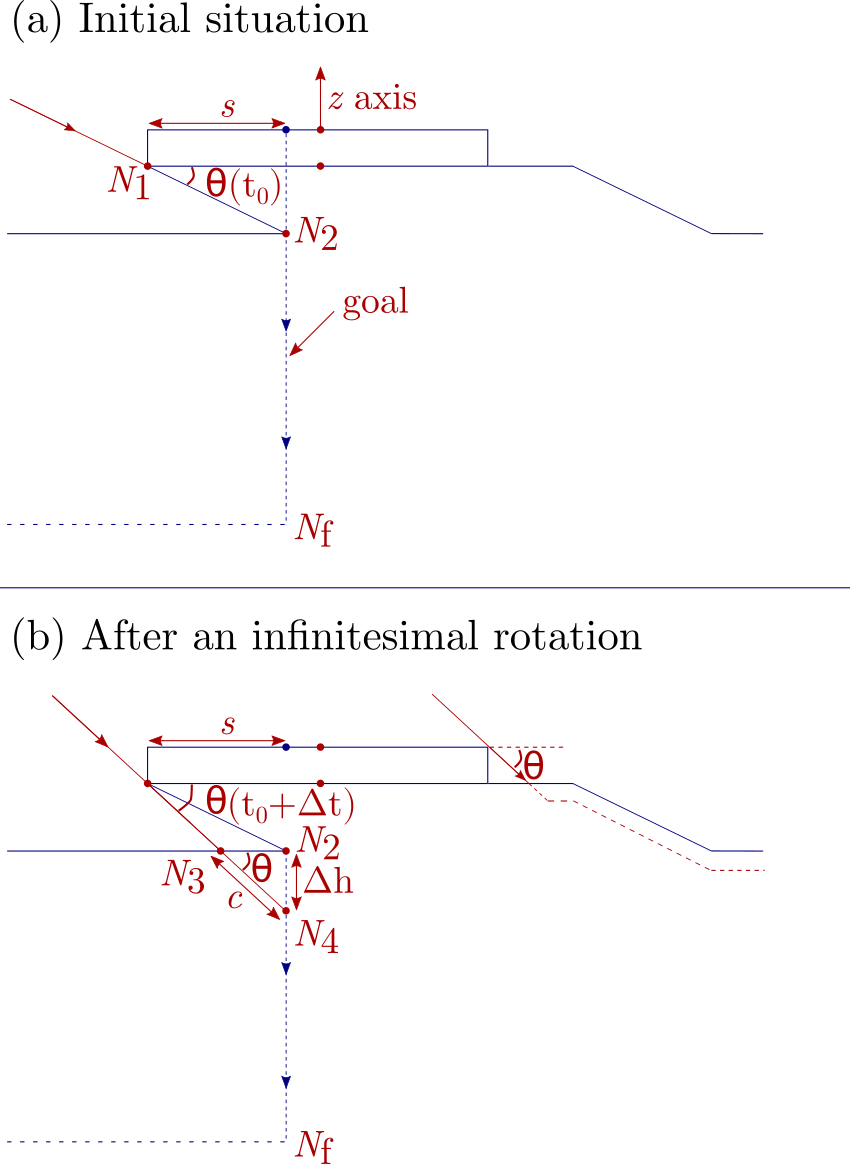}
			\caption{An angled dry etch during an infinitesimal rotation. Two different angles of penetration, $\theta\left(t_0\right)$, and $\theta\left(t_0 + \Delta t \right)$, are shown for corresponding times $t = t_0$ and $t = t_0 + \Delta t$, respectively. (a) The initial angle at which the fixed-angle etch progresses from the point $N_1$ to the target depth $N_2$. (b) The new, modified trajectory of the angled etch after an infinitesimal rotation is applied. The dry etch now makes impact at point $N_3$, and the point $N_2$ is shielded by the protective resin.  Both (a) and (b) are simplified schematics that assume an infinite selectivity of the etch. It is assumed none of the protective resin is removed during the infinitesimal time $\Delta t$.}
			\label{Time-depEtch}
			\end{figure}

			This method eliminates the initial step of micro-fabrication of starting structures. The process can be summarized as two steps:
			\begin{enumerate}	
				\item Time-dependant angled dry etch of a flat surface with a resin pattern
				
				\item Wet etching to further reduce the size of the support structure
			\end{enumerate}
			By varying the angle of penetration during the dry etch, one can generate similar structures as using Method 1, but starting from a flat surface with a pattern of resin deposited on its surface. To achieve this, assuming an infinite selectivity ratio $k$, letting $r$ be the etching rate in meters per second, and with $\theta$ and $s$ the same as in Method \ref{PreFabMicroStruct}, the angle of incidence must vary in time according to
			\begin{equation}\label{ThetaOft}
			\frac{d \theta\left(t\right)}{dt} = \frac{r \left( \theta\left(t\right) \right) }{s}\sin{\left(\theta\left(t\right)\right)}\cos^2{\left(\theta\left(t\right)\right)}.
			\end{equation}
			We will now prove equation \eqref{ThetaOft}. Consider Fig. \ref{Time-depEtch} (a) as the initial condition, where an etch at a fixed angle has brought the near-point of etching contact $N_1$ a horizontal distance $s$ away from the edge of the resin, to $N_2$. From this configuration, the goal is to produce a vertical column. To accomplish this, the angle of penetration $\theta$ of the dry etch can be gradually increased at just the right rate to produce a straight vertical line perpendicular to the plane of the surface. The goal is to find an expression for this rate $d \theta\left(t\right) / dt$.
			We consider an infinitesimal increase in the angle of penetration from $\theta \left( t_0 \right)$ to $\theta = \theta\left(t_0 +\Delta t\right)$, in an infinitesimal time step $\Delta t$, as shown in Fig. \ref{Time-depEtch} (b). This moves the shadow cast by the mask horizontally to the left along the flat surface, and displaces the near-point of etching contact $N_2$ to the point $N_3$. In this same infinitesimal time step $\Delta t$, the etching must progress a diagonal distance $c = r \Delta t$ to reach the vertical target line, where $r$ is the etching rate in meters per second. This forms the hypotenuse of a small triangle. During this time, the vertical distance etched along the axis of the target line is $\Delta h = \left(\Delta h/\Delta t\right) \times \Delta t$, where $\left(\Delta h/\Delta t\right)$ is the instantaneous etching velocity in the vertical direction. The time step $\Delta t$ is the same in the expressions for $c$ and $\Delta h$, so dividing the expression for $\Delta h$ by the expression for $c$ gives $\Delta h /c = \left(\Delta h/\Delta t\right) \times \Delta t / r \Delta t~$, which rearranges to
			\begin{equation}\label{Vert_EtchVel}
			\left( \Delta h/\Delta t \right) = \Delta h r/c = \sin{\theta\left(t\right)r},
			\end{equation}
			where we have defined $\theta\left(t_0 +\Delta t\right) \equiv \theta\left(t\right)$. One can observe that equation \eqref{Vert_EtchVel} is the $z$-component of the etching velocity, if we let $\vec{r}$ denote the etch rate and direction of penetration of the etching agent, such that $\Delta h/\Delta t = \vec{v}_z = -\left( \vec{r} \cdot \hat{z} \right) \hat{z} = -r \sin{\theta\left(t\right)} \hat{z}$. Equation \eqref{Vert_EtchVel} sets the rate at which the shadow cast by the mask must move along the target line. The height of the shadow on the target line is $h = s\tan{\theta}$,~ so taking the derivative with respect to time gives 
			\begin{equation}\label{ShadowMovement}
			dh/dt = \frac{\partial h}{\partial \theta} \frac{d\theta}{dt} = \frac{s}{\cos^2{\theta\left(t\right)}}\frac{d\theta\left(t\right)}{dt}.
			\end{equation}
			Imposing that $\Delta h/\Delta t$ in equation \eqref{Vert_EtchVel} must equal $dh/dt$ in equation \eqref{ShadowMovement} and rearranging leads to
			\begin{equation}\label{ThetaOft2}
			\frac{d\theta\left(t\right)}{dt} = \frac{r}{s}\sin{\theta\left(t\right)}\cos^2{\theta\left(t\right)},
			\end{equation} 
			and this is equation \eqref{ThetaOft}. For an angle-dependent etch rate the derivation remains the same, but the fixed etch rate $r$ can be replaced by an angle-dependent expression $r \left( \theta \left( t \right) \right)$. If furthermore, the dry etch is anisotropic such that the angle of incidence of the dry etch is $\theta ' = f(\theta)$, it is useful to know the equivalent to expression \eqref{ThetaOft2} in terms of the attack angle $\theta '$. In other words, one would like to know the expression for $d\theta ' \left(t\right) / dt$ in terms of $\theta '$. This can be obtained if the relationship $\theta ' = f\left( \theta \right)$ is known. In general,
			\begin{equation}	
			\frac{d\theta ' \left(t\right)}{dt} = \frac{d}{dt} f \left( \theta \right) = \frac{\partial  f \left( \theta \right)}{\partial \theta} \times \frac{d\theta \left(t\right)}{dt},
			\end{equation}
			so $d\theta ' \left(t\right) / dt$ can be obtained by multiplying the right hand side of equation \eqref{ThetaOft} or \eqref{ThetaOft2} by the derivative of $f \left( \theta \right)$ with respect to $\theta$. Once the rate of the change of the attack angle $d\theta ' \left(t\right) / dt$ is expressed in terms of the instantaneous penetration angle $\theta$, this equation can be rewritten exclusively in terms of the attack angle $\theta '$ by plugging in $\theta = f ^{-1} \left( \theta ' \right)$.

			\begin{figure}[h]
			\centering
			\includegraphics[width = 0.5\textwidth]{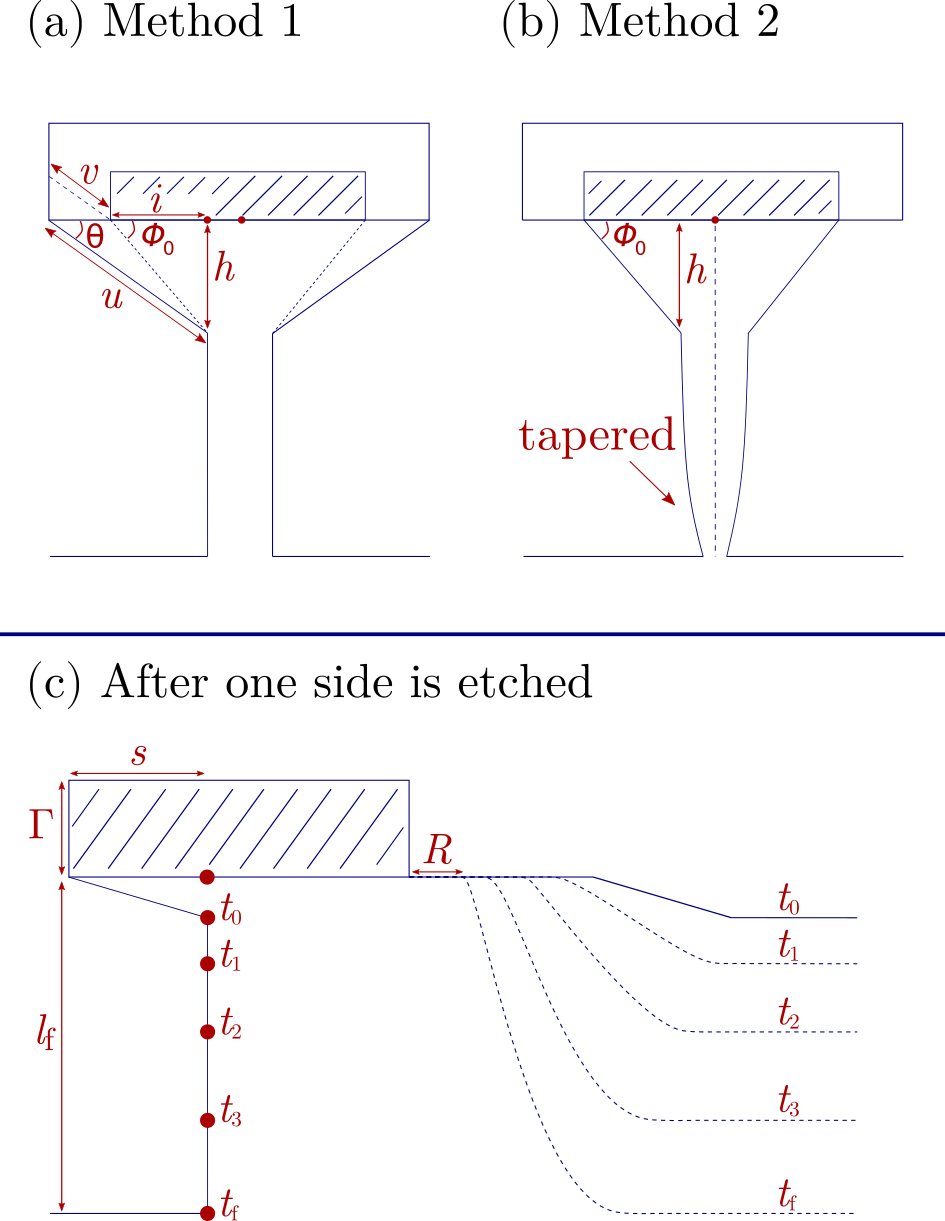}
			\caption{Top: the effect of non-infinite selectivity on the results of Method 1 (left) and Method 2 (right) after angled dry etches. (a) A fixed-angle dry etch results in a support column with uniform thickness from top to bottom. (b) An uncompensated time-dependent angled dry etch results in a tapered support column. Bottom: the shadow cast by the resin during a time-dependent angled etch incident from the left, carves away a sigmoid shape on the lee side of the resin. The shape of the sigmoid is shown at five different times during the etching process.}
			\label{ResultMethod1vs2}
			\end{figure}		
			For a non-infinite selectivity, the angled etch removes some of the protective resin (see Fig. \ref{DryAndWetEtch} (a) and associated discussion). This means the height of the shadow cast by the resin is smaller than the value used for the calculations, and the vertical movement of the shadow along the target line becomes slower than calculated. These effects have two implications. First, the fact that less material is protected by the the resin's shadow means that the initial angle used at the start of the time-dependent etch should not be $\theta_0$~, but rather $\phi_0 = \arctan{\left(h/i\right)}$, where the angle $\phi_0$ is shown in Figs. \ref{DryAndWetEtch} (a) and \ref{ResultMethod1vs2} (a), and $i$ is the distance of the overhang after the initial angled etch. Second, since additional resin is removed over the course of the time-dependent etch, lower parts of the pillar receive more exposure than calculated, causing the pillar to taper towards the bottom. The angles at the base of the pillar become $\le 90^\mathrm{o}$, with exact values depending on the etching selectivity $k$ (Fig. \ref{ResultMethod1vs2} (b)). Since the amount of tapering is difficult to estimate, a pragmatic approach is to obtain images of the pillars after the time dependent dry etch and before the wet etch. This allows the wet etch to be adjusted to avoid completely dissolving the tapered portion of the support pillar. For a fine-tuned result it may be desirable to perform microscopy at intermediate times during the time-dependent dry etch and observe how the profile of the support pillar evolves. The rate of change of the angle, $d \theta\left(t\right) / dt$, of the time-dependent etch can then be adjusted (increased) based on a curve generated from various benchmarked times and associated tapering depths, to compensate for tapering. Figure \ref{ResultMethod1vs2} (a) and (b) show a comparison of possible final shapes resulting from Method 1 and Method 2 for non-zero etching selectivities.
			
			In Method 2, the angled dry etch incident from the left removes material both on the left side of the resin, and on the right side of the resin, as shown in Figure \ref{ResultMethod1vs2} (c). In particular, the time-dependent angled etch traces an evolving sigmoid curve on the right side of the resin. The profile of this sigmoid curve influences how etching must be performed after the sample is rotated by $180^{\mathrm{o}}$. Letting $\Gamma$ be the height of the mask and $l_\mathrm{f}$ be the final height of the pillar after etching from the left side, we can identify two limiting cases: $\Gamma \gg l_\mathrm{f}$, and $\Gamma \ll l_\mathrm{f}$. If etching is performed in the first regime, the distance $R$ between the edge of the mask and the shoulder of the sigmoid curve remains large compared to $l_\mathrm{f}$, and the second etch should be time-dependent, according to equation \eqref{ThetaOft}. However, when $\Gamma \ll l_\mathrm{f}$ the distance $R$ tends to zero and the second etch should use a fixed angle, as described in Method 1. The more convenient regime will largely be determined by the selectivity of the etch. For high selectivities, a thinner mask can be used allowing the second etch to be done at a fixed angle. On the other hand, for low selectivities where a thin mask is not suitable, one way to still achieve a predictable outcome is to significantly increase the mask height, leading to the $\Gamma \gg l_\mathrm{f}$ regime. Regardless of whether one uses the $\Gamma \gg l_\mathrm{f}$ or the $\Gamma \ll l_\mathrm{f}$ regime, the etches at rotations of $90^{\mathrm{o}}$ and $270^{\mathrm{o}}$ must be performed using fixed angle etches. This is because these sides are not shielded by the square cap of protective resin; the first time-dependent etch removes material up to a depth $l_f$ both on the $90^{\mathrm{o}}$ and $270^{\mathrm{o}}$ sides. If one is in the $\Gamma \ll l_\mathrm{f}$ regime, one way to avoid etching the plane of the sample too deeply is by depositing a protective layer at normal incidence ($\theta_0 = 90^{\mathrm{o}}$) after the first etch. This ensures that the subsequent fixed-angle etches at rotations of $180^{\mathrm{o}}$, $90^{\mathrm{o}}$ and $270^{\mathrm{o}}$ around the $z-$axis do not erode the plane of the sample, which would increase the relative height of the pillar (for example, to $4 \times l_\mathrm{f}$ if the time-dependent angled etch and the fixed-angle etches erode the same vertical distance). If one uses this method of depositing a protective layer in the $\Gamma \gg l_\mathrm{f}$ regime, it is necessary to wait to deposit the protective layer until after the second etch at a rotation of $180^{\mathrm{o}}$. In this case, the second etch increases the height of the pillar above the surrounding plane by an additional amount $l_\mathrm{f}$. The extra height must be taken into account in the initial dimensioning of the system. For both the $\Gamma \gg l_\mathrm{f}$ and the $\Gamma \ll l_\mathrm{f}$ regime, the deposition of a protective layer at normal incidence results in some protective material coating the sides of the pillar. This layer is thin compared to the coating in the plane of the sample. Nevertheless, when fixed-angle etches are performed after deposition of the protective layer, the duration must be increased to account for removal of the protective coating on the sides of the pillar. 
			
			Once a combination of time-dependent and fixed-angle dry etches have been used to create an upside-down pyramid on top of a support pillar, a wet etch can further reduce the thickness of the support pillar. For dimensioning purposes, we now consider the final dimensions of the top plateau when Method 2 is applied. Many results from Method 1 can be reused. The fixed angle etch at an angle $\theta_0$ in Method 1 has exactly the same effect on the top plateau as the fixed angle etch at an angle $\theta_0$ in Method 2. Similarly, the wet etch in Method 1 has the same effect as the wet etch in Method 2. Therefore, to a first approximation the final dimension of the top plateau is again given by:
			\begin{equation}\label{Method2r_f}
			r_\mathrm{f} = s \left( 1 - 1/k \right) + (e + b) - b / \sin{\phi_0}.
			\end{equation}
			However, for non-infinite selectivities the time-dependent etch further reduces the dimension of the top plateau, because it removes more protective resin and more of the top plateau is exposed. Assuming this produces an additional reduction of $X~\mu$m, the adjusted final dimension, which can be denoted $r^{\mathrm{t}}_\mathrm{f}$, is given by: 
			\begin{equation}\label{TimeDep_r_f}
			r_\mathrm{f} \ge r^{\mathrm{t}}_\mathrm{f} = s \left( 1 - 1/k \right) + (e + b) - b / \sin{\phi_0} - X.
			\end{equation}
			To go from equation \eqref{Method2r_f} to equation \eqref{TimeDep_r_f} we let $s = u \cos{\theta_0}$ and all other parameters remain fixed, and observed that the time-dependent etch reduces the final half side-length to  $r^{\mathrm{t}}_{\mathrm{f}}$. For design purposes, it is more useful to fix the target half side-length $r_{\mathrm{f}}$ and see how the time-dependent etch affects the dimension $s$ (or $u \cos{\theta_0}$, or $u \cos{ \big( \tan^{-1} \left( (1-1/k) \tan ( \phi_0 ) \right) \big) }$   ), of the initial protective resin. Using equation \eqref{Method2r_f} but now fixing the target dimension $r_\mathrm{f}$ and solving for $u$, we find the depth $u$ of the initial fixed-angle etch must be increased to $u^{\mathrm{t}}$:
			\begin{eqnarray}\label{u-time-dep_convenient}
			&& u \le u^{\mathrm{t}} = \left[b \left(\frac{1}{\sin{\phi_0}} - 1 \right) + r_f - e + X \right] 
			\nonumber \\
			&& \times \frac{1}{\cos \left(
				\tan^{-1} \left( \left( 1 - 1/k \right) \tan \phi_0 \right) \right) \left( 1 - 1/k \right)}.
			\end{eqnarray}
			From this, and letting $r^{\mathrm{t}}_\mathrm{i}$ denote the initial half side-length of the protective resin when a time-dependent etch is used, the side length of the initial square of protective resin which should be deposited is given by $2r^{\mathrm{t}}_\mathrm{i} = 2 \times \left( u^{\mathrm{t}} \cos{\theta_0} + w \right)\mu$m.
			\newline			
			\section*{Conclusions}
			We have outlined two approaches for creating large overhangs using a combination of angled dry ballistic RIE etching and wet etching. Both approaches are suitable for designing specialized structures at the micrometer and nanometer scale. In particular, we illustrate the principles using a theoretical example for making electrodes with dimensions on the order of $50~\mu$m, perched atop $5~\mu$m diameter, low-capacitance conducting pillars $150~\mu$m to $200~\mu$m tall. This example is targeted towards the design of future quantum computing systems.

			\section*{Acknowledgments}
			The authors would like to thank Yaping Ren from the Center for Quantum Technologies and Nicolas Reyren from the CNRS/Thales UMR137 for helpful conversations and technical insights. This work is supported by a Singapore Ministry of Education Academic Research Fund Tier 2 grant (No. MOE2016-T2-2-120) and an NRF-CRP grant (No. NRF-QEP-P6). N. Van Horne acknowledges the CQT Ph.D. scholarship from the Centre for Quantum Technologies at the National University of Singapore.

\bibliography{References}

\end{document}